\documentclass[twocolumn,showpacs,preprintnumbers,amsmath,amssymb]{revtex4}
\usepackage{graphicx}% Include figure files
\usepackage{dcolumn}% Align table columns on decimal point
\usepackage{bm}% bold math

\begin{document}

%\preprint{APS/123-QED}

\title{Stimulated emission of two photons in parametric amplification\\
and its interpretation as multi-photon interference}
% Force line breaks with \\

\author{F. W. Sun$^{1}$}
\author{B. H. Liu$^{1}$}
\author{Y. X. Gong$^{1}$}
\author{Y. F. Huang$^{1}$}
\author{Z. Y. Ou$^{2,1}$}
\email{zou@iupui.edu}
\author{G. C. Guo$^{1}$}
 \affiliation{$^1$Key Laboratory of Quantum Information,
 University of Science and Technology of China,\\ CAS, Hefei, 230026, the People's Republic of China
 \\$^2$Department of Physics, Indiana
University-Purdue University Indianapolis \\ 402 N. Blackford
Street, Indianapolis, IN 46202}%Lines break automatically or can be forced with \\

\date{\today}% It is always \today, today,

\begin{abstract}
Stimulated emission of two photons is observed experimentally in
the parametric amplification process and is compared to a
three-photon interference scheme. We find that the underlying
physics of stimulated emission is simply the constructive
interference due to photon indistinguishability. So the observed
signal enhancement upon the input of photons is a result of
multi-photon interference of the input photons and the otherwise
spontaneously emitted photon from the amplifier.

\end{abstract}

\maketitle

Stimulated emission, first proposed by Einstein \cite{ein} to
explain the blackbody radiation spectrum, is the main process in
laser operation. It provides the optical gain of an active medium
and is responsible for the coherence of laser light \cite{eber}.
Although the process was studied extensively as an amplification
process of a classical wave field as early as in 1955 \cite{town},
its effect on the nonclassical state of light was only
investigated not long ago \cite{man}, especially in the contest of
quantum state cloning \cite{zei,bouw,how}.

Fundamentally, stimulated emission occurs at single-photon level,
i.e., it is seen as the creation of an identical photon to an
incoming photon. However, the same photon can also be produced
even without the incoming photon, due to spontaneous emission.
Thus, the existence of the input photon will enhance the
production rate, as compared to the case without the input photon.
Indeed, in recent study of stimulated emission by single photons,
a doubled rate is observed in photon production that is correlated
to the input photon \cite{ou,bouw,how}. But the above picture is
only phenomenological and does not tell the underlying physics. So
what fundamental physical principle governs this phenomenon?

If we make a detailed analysis of the enhancement due to
stimulated emission, we find that it bears some resemblance to the
famous Hanbury Brown-Twiss effect \cite{hbt}, i.e., the photon
bunching effect of thermal light \cite{ou,orw2}: the enhancement
factor in both cases is one-fold and the temporal profile is the
same. The photon bunching effect, as pointed out by Glauber
\cite{glau2} in 1965, is in essence a two-photon interference
effect. This suggests that the underlying physics in stimulated
emission is simply multi-photon quantum interference. Recently,
Khan and Howell \cite{how2} and Irvine {\it et al.} \cite{irv}
utilized a beam splitter and the two-photon Hong-Ou-Mandel
interference effect \cite{hom,rar} to emulate the photon cloning
process observed in stimulated emission \cite{bouw}. This further
demonstrates the connection between stimulated emission and the
multi-photon interference.

\begin{figure}
\includegraphics[width = 3in]{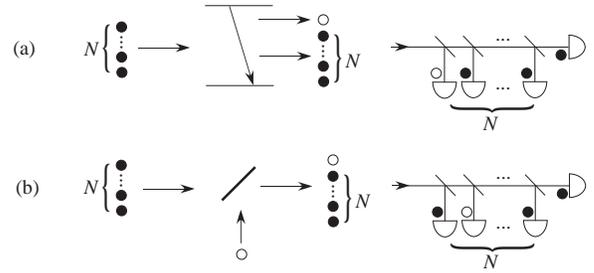}
\caption{Comparison between (a) the stimulated emission and (b)
the multi-photon interference. }
\end{figure}

In this letter, we wish to report on an experiment in support of
the above claim. In the experiment, we inject a two-photon state
into a parametric amplifier and observe an enhancement by a factor
of nearly two in the photon production that is correlated to the
input photons. Furthermore, we mimic the same phenomenon with a
beam splitter, in a similar manor to Ref.\cite{rar} for the
single-photon input case. These two phenomena can be viewed as a
generalized three-photon bunching effect and are a result of
three-photon constructive interference due to photon
indistinguishability.

To understand the connection between the stimulated emission and
multi-photon interference, we consider the two situations in
Fig.1. The process of stimulated emission of an $N$-photon state
is shown in Fig.1(a), where $N$ photons interact with an atom in
an excited state. The atom will emit one photon regardless of the
input. Total photon number is $N+1$. Assume that the spontaneous
emission rate is $R$ into the same mode of the input photons. It
is known that the emission rate stimulated by one photon is the
same. Then since each input photon may stimulate the excited atom,
the total rate is then $(N+1)R$. The extra $R$ is from spontaneous
emission. There is an enhancement factor of $N$ in the photon
production rate. The case of $N=1$ was observed in
Ref.\cite{bouw}.

In multi-photon interference in Fig.1(b), on the other hand, a
single photon and $N$ photons are combined by a 50:50 beam
splitter. The probability of detecting all $N+1$ photons in one
side is easily calculated to be $(N+1)/2^{N+1}$ (see below).
However, when the single photon is distinguishable from the $N$
photons and no multi-photon interference occurs, we find the
detection probability is simply $1/2^{N+1}$. The $N+1$ factor is a
result of constructive interference of $N+1$ possibilities in
detecting the $N+1$ photons. Each possibility corresponds to the
situation when the single input photon is detected by a specific
detector. (Fig.1 shows two such possibilities in the $N+1$-photon
detection.) We add the amplitudes of the $N+1$ possibilities
before taking the absolute value for the indistinguishable case
but we add the absolute values of the amplitude of each
possibility for the distinguishable case. The ratio between the
two cases is then $N+1$.  The case of $N=1$ is the Hong-Ou-Mandel
photon bunching effect \cite{rar}. A similar interference effect
was observed by Ou {\it et al.} \cite{orw} with a $|2,2\rangle$
input state. Notice that the enhancement factor here is the same
as the stimulated emission in Fig.1(a). Therefore, the spontaneous
emission rate $R$ corresponds to the situation when the input $N$
photons to the atom are distinguishable from the emitted photon by
the atom, so that the atom is not influenced by the input photons
and only emits spontaneously. This case is exactly the same as the
case in Fig.1(b) but when the single photon is distinguishable
from the $N$ photons.  When the input $N$ photons are
indistinguishable from the emitted photon by the atom,
constructive multi-photon interference leads to a factor of $N+1$
enhancement in photon detection rate. Thus, the underlying physics
in the stimulated emission is the photon indistinguishability that
results in multi-photon interference.

In the following, we will use parametric amplifier to study the
stimulated emission by multiple photons. Mathematically, any phase
insensitive linear amplifier can be modelled as a single mode
parametric amplifier, which is described quantum mechanically by
\cite{cav}
\begin{eqnarray}
\hat a_s^{(out)} = G\hat a_s +g \hat a_i^{\dag}, \label{1}
\end{eqnarray}
where $\hat a_i$ corresponds to the internal modes of the
amplifier and is the idler mode for the parametric amplifier. It
is usually independent of $\hat a_s$ and is in vacuum. To preserve
the commutation relation, we need $|G|^2 - |g|^2 =1$.

At microscopic level of atoms, we have a small value of $|g|<<1$
or $|G|\sim 1$. The unitary evolution operator for Eq.(\ref{1})
then has the form of
\begin{eqnarray}
\hat U \approx 1 + (g \hat a_s^{\dag}\hat a_i^{\dag} +
h.c.)\label{2}
\end{eqnarray}
With a vacuum input of $|0\rangle$, we have the output state
\begin{eqnarray}
|\Phi\rangle_{out}^{(0)} =\hat U |0\rangle  \approx |0\rangle + g
|1_s\rangle\otimes|1_i\rangle.\label{3}
\end{eqnarray}
The last term gives the spontaneous emission with a probability
of $|g|^2$. When the input is a single-photon state
$|1_s\rangle\otimes|0_i\rangle$, we have
\begin{eqnarray}
|\Phi\rangle_{out}^{(1)} &\approx & |1_s\rangle|0_i\rangle + g
(\hat a_s^{\dag}|1_s\rangle)\otimes(\hat a_i^{\dag}|0_i\rangle
)\cr &= &|1_s\rangle|0_i\rangle + \sqrt{2}g
|2_s\rangle\otimes|1_i\rangle.\label{4}
\end{eqnarray}
The probability for the emission from the amplifier is then
$2|g|^2$. The extra emission probability of $|g|^2$ is usually
attributed to the stimulated emission, which is similar to the
photon bunching effect \cite{hbt,ou,orw2,rar}.

When the input state is a two-photon state of
$|2_s\rangle|0_i\rangle$, we have the output state as
\begin{eqnarray}
|\Phi\rangle_{out}^{(2)} &\approx & |2_s\rangle|0_i\rangle + g
(\hat a_s^{\dag}|2_s\rangle)\otimes(\hat a_i^{\dag}|0_i\rangle )
\cr &= &|2_s\rangle|0_i\rangle + \sqrt{3}g
|3_s\rangle\otimes|1_i\rangle.\label{5}
\end{eqnarray}
The photon emission rate from the amplifier is now three times the
spontaneous rate. In fact, it is straightforward to find that,
with an $N$-photon state as the input, the rate of photon emission
from the amplifier is $N+1$ times the spontaneous emission rate.
As the single-photon input case, each fold of increase in the rate
can be attributed to the stimulated emission from one individual
photon in the input $N$-photon state.

Notice that when the input photons are not in the same mode as the
amplifier and thus are distinguishable from the photon emitted by
the amplifier, the output state becomes
\begin{eqnarray}
|\Phi\rangle_{out}^{(N)'} &\approx &
|0_s\rangle|N_{s'}\rangle|0_i\rangle + g (\hat
a_s^{\dag}|0_s\rangle) \otimes|N_{s'}\rangle\otimes(\hat
a_i^{\dag}|0_i\rangle ) \cr &=
&|0_s\rangle|N_{s'}\rangle|0_i\rangle + g
|1_s\rangle|N_{s'}\rangle |1_i\rangle, \label{6}
\end{eqnarray}
where $N \ge 1$. So the photon production rate is exactly the same
as the spontaneous emission.

The above analysis with the parametric amplifier confirms the
previous results obtained from the pictorial argument based on
Fig.1. Next, we will demonstrate experimentally the enhancement
effect for a two-photon input and compare it with a three-photon
interference scheme with a beam splitter.

\begin{figure}
\includegraphics[width = 2.9in]{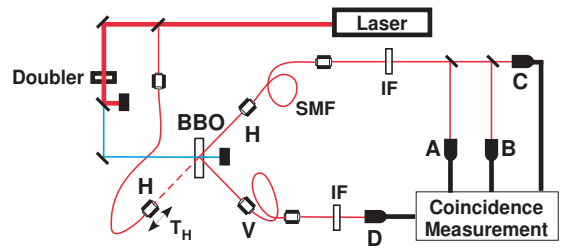}
\caption{Schematics for studying the stimulated emission of an
input of $N$-photon state with parametric amplification. SMF:
single-mode fiber; IF: interference filter; H,V: horizontal and
vertical polarizations; $T_H$: adjustable delay.}
\end{figure}

The experimental arrangement for studying the stimulated emission
is sketched in Fig.2. A mode-locked Ti:sapphire laser operating at
780 nm is frequency doubled and the harmonic field serves as the
pump field for a parametric amplifier made of a 1-mm long
$\beta-$Barium Borate (BBO) crystal. The crystal is so oriented
that it satisfies the type-II phase matching condition and
beam-like fields are generated \cite{tsu}. A small portion is
split from the Ti:sapphire laser and serves as the input field to
the signal port of the parametric amplifier. The injected coherent
field is heavily attenuated down to a rate much less than one
photon per pulse. But even so, the coherent state consists of
vacuum, one-photon state, two-photon state, and more. So the
output state is a superposition of the states in
Eqs.(\ref{3}-\ref{5}). Therefore, in order to observe the
enhancement effect in stimulated emission by a specific number of
photons, we need to make a projection measurement to the
corresponding states in Eqs.(\ref{4}, \ref{5}). For example, for a
two-photon state input, the projection is to the second term in
Eq.(\ref{5}). This is achieved by a four-photon coincidence
measurement, as depicted in Fig.2. Joint measurement with the
idler photon is necessary to discriminate against the three-photon
contribution directly from the injected coherent field. Photon
(in)distinguishability between the input photons and the photon
emitted from parametric down-conversion is realized by an
adjustable delay $T_H$ on the coherent injection field. A
single-mode fiber (SMF) is used to collect the signal field from
the amplifier, in order to ensure a good spatial mode match.
Interference filters of bandpass of 3 $nm$ are used for temporal
mode cleaning.

\begin{figure}
\includegraphics[width = 2.8in]{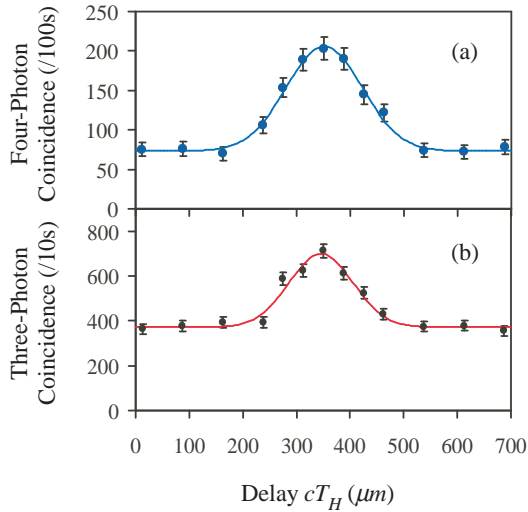}
\caption{(a) Four-photon coincidence of ABCD detectors in 100s and
(b) three-photon coincidence of ABD detectors in 10s as a function
of the delay $cT_H$.}
\end{figure}

The conditions for the situations in Eqs.(\ref{5}, \ref{6}) are
satisfied by adjusting the delay $T_H$. When the delay is right,
the injected coherent field pulse arrives in time with the pump
pulse to the amplifier and the photon emitted by the amplifier is
indistinguishable from the incoming photons in the coherent state.
But when the delay is either too large or too small, there is no
overlap between the coherent pulse and the pump pulse. This is the
situation described in Eq.(\ref{6}). Therefore, as we scan the
delay $T_H$, the four-photon coincidence of A, B, C, D detectors
should exhibit a bunching effect with a peak-to-wing ratio close
to three. Fig.3(a) shows the result of the measurement. The error
bars are the statistical errors of one standard deviation. The
solid curve is a least square fit of the data to a Gaussian of the
form
\begin{eqnarray}
F(T_H) = A\Big[1+v e^{-(T_H-T_0)^2/T_c^2}\Big],
\end{eqnarray}
where $T_0$ is the center position of the peak and $T_c$ is
related to the width of the peak. We obtain $v_2=1.81 \pm 0.15$ as
the enhancement factor for the data in Fig.3(a), which gives 2.81
as the ratio between the peak and the wing. This value is close to
the ideal value of three in Eq.(\ref{5}) for the stimulated
emission by two photons.

In the meantime, three-photon coincidences of ABD detectors are
also registered and are shown in Fig.3(b). This measurement
corresponds to the second term in Eq.(\ref{4}) and gives the
stimulated emission by one input photon. The Gaussian fit gives an
enhancement factor of $v_1=0.88\pm0.14$. The peak-to-wing ratio of
1.88 is close to the ideal value of two in Eq.(\ref{4}).

The reason for the imperfection in the experiment is due to mode
mismatch between the input field and the amplifier mode, {\it
i.e.}, mismatch between $|N\rangle$ and the mode for which the
operator $\hat a_s$ represents. Although the spatial mode is
matched by the single-mode fiber (SMF in Fig.2), the temporal mode
is hard to match because the temporal coherence of the parametric
down-conversion process is very complicated and the fields are not
transform-limited even if the pump field is. Nevertheless, we use
interference filters to clean up the temporal profile. The full
widths of the peaks in Fig.3 is approximately $2T_c= 660ps$, close
to the coherence time of the interference filters (IF in Fig.2) of
width 3 $nm$.

Next we consider the situation depicted in Fig.1(b) where an
$N$-photon state is superposed with a single-photon state by a
50:50 beam splitter. The output state for the beam splitter is
given by \cite{cam}
\begin{eqnarray}
|\Phi\rangle_{out}^{(BS)} = \sqrt{N+1\over
2^{N+1}}|N+1\rangle_1|0\rangle_2 + ...~,\label{8}
\end{eqnarray}
where we only write down the state for which all the $N+1$ photons
exit at one port (port 1) of the beam splitter. On the other hand,
if the input $N$ photons are distinguishable from the single
photon from the other input port, they behave like classical
particles and follow the Bernoulli distribution. The output state
becomes
\begin{eqnarray}
|\Phi\rangle_{out}^{(BS)'} = \sqrt{1\over
2^{N+1}}|N\rangle_{1'}|1\rangle_1|0\rangle_2 + ...~.\label{9}
\end{eqnarray}
Therefore, the rate of detecting $N+1$ photons in port 1 is $N$
times bigger when the photons are all indistinguishable than when
the $N$ photons are distinguishable from the one photon. As
discussed before, this increase stems from a constructive
$N+1$-photon interference.

\begin{figure}
\includegraphics[width = 2.9in]{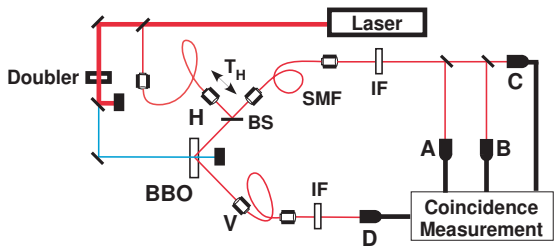}
\caption{A simple scheme for interference of $N$ photons and one
photon with a beam splitter. Same notations as Fig.2.}
\end{figure}

\begin{figure}
\includegraphics[width = 2.9in]{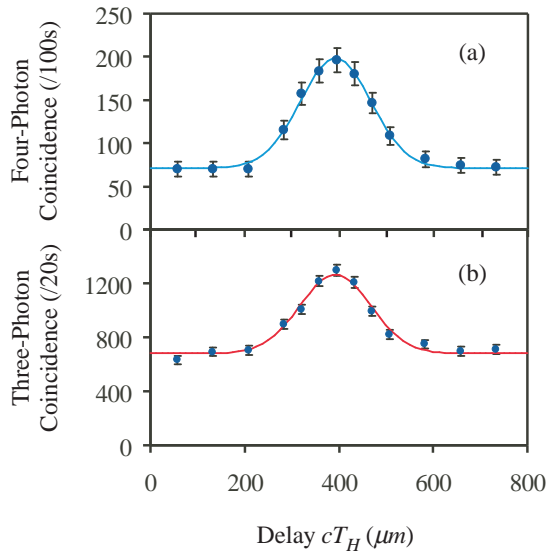}
\caption{Same as Fig.2 but with data obtained from Fig.4.}
\end{figure}

The experimental arrangement with a beam splitter is similar to
Fig.2 and is shown in Fig.4, where the split weak coherent field
is directed to a beam splitter to combine with the signal field
(H) from the parametric down-conversion. In order to mimic the
stimulated emission process shown in Fig.2, all the experimental
parameters such as pump power, the strength of coherent field,
etc. are the same as those in Fig.2. We adjust the delay $T_H$ on
the coherent field to ensure the temporal overlap between the
coherent field and the down-converted photon. When gated on the
detection of the V-photon by detector D, the H-field of the
down-conversion is in a single-photon state. But because of the
complicated dispersion in the parametric down-conversion process,
the single-photon state is not transform-limited. Again,
interference filters are used to clean up some of the temporal
modes but impossible to achieve the perfect match.

We record both the four-photon coincidence of ABCD detectors and
the three-photon coincidence of ABD detectors. The former
corresponds to the $N=2$ case in Eq.(\ref{9}), whereas the latter
to the $N=1$ case. The results are shown in Fig.5. The fitted
curves are very similar to Fig.3 but with $v_2= 1.78\pm 0.14$ and
$v_1 =0.86\pm0.09$.

As can be seen, Figs.3 and 5 show the same result within the
statistical errors. This confirms our claim that the underlying
physics in stimulated emission is nothing but multi-photon
interference. The interference effect is a result of
indistinguishability between the input photons and the photon
emitted by the amplifier.

\begin{acknowledgments}
This work was funded by National Fundamental Research Program of
China, the Innovation Funds of Chinese Academy of Sciences. ZYO is
supported by the US National Science Foundation under Grant No.
0245421 and No. 0427647.
\end{acknowledgments}


\begin{thebibliography}{99}
\bibitem{ein} A. Einstein, Phys. Z. {\bf 18}, 121 (1917).

\bibitem{eber} P. W. Milonni and J. H. Eberly, {\it Lasers} (Wiley, New York, N.Y., 1988).

\bibitem{town} J. P. Gordon, H. J. Zeiger, and C. H. Townes, Phys. Rev. {\bf 99},
1264 (1955).

\bibitem {man} S. Friberg and L. Mandel, Opt. Commun. {\bf 46} 141
(1983).

\bibitem {zei} C. Simon, G. Weihs, and A. Zeilinger,
Phys. Rev. Lett. {\bf 84}, 2993 (2000).

\bibitem {bouw}  A. Lamas-Linares,  J. C. Howell, and  D.  Bouwmeester,   Nature, {\bf
412} 6850 (2001).

\bibitem {how}  A. Lamas-Linares, C. Simon, J. C. Howell, and  D.  Bouwmeester, Science {\bf 296}, 712
(2002).

\bibitem {ou} Z. Y. Ou, L. J. Wang, and L. Mandel, J. Opt. Soc. Am. B {\bf 7}, 211 (1990).

\bibitem {hbt} R. Hanbury-Brown and R. W. Twiss, Nature {\bf 177}, 27
(1956).

\bibitem {orw2} Z. Y. Ou, J.-K. Rhee, and L. J. Wang,  Phys. Rev. A {\bf 60},
593 (1999).

\bibitem {glau2} R. J. Glauber, in {\it Quantum Optics and Electronics
(Les Houches Lectures)}, p.63, edited by C. deWitt, A. Blandin,
and C. Cohen-Tannoudji (Gordon and Breach, New York, 1965).

\bibitem {how2} I. A. Khan and J. C. Howell, \pra {\bf 70}, 010303(R)
(2004).

\bibitem {irv} W. T. M. Irvine, A. Lamas-Linares, M. J. A. de Dood, and D. Bouwmeester, \prl {\bf 92},
047902 (2004).

\bibitem{hom} C. K. Hong, Z. Y. Ou, and L. Mandel, \prl {\bf 59}, 2044 (1987).

\bibitem {rar} J. G. Rarity and P. R. Tapster, J. Opt. Soc. Am. B {\bf 6},
1221 (1989).

\bibitem{orw} Z. Y. Ou, J.-K. Rhee, and L. J. Wang, Phys. Rev. Lett. {\bf 83}, 959
(1999).


\bibitem{cav} C. M. Caves, Phys. Rev. D {\bf 26}, 1817
(1982).

\bibitem{tsu} S. Takeuchi,  Opt. Lett. {\bf 26}, 843 (2001).

\bibitem{cam} R. A. Campos, B. E. A. Saleh, and M. C. Teich, Phys. Rev. A{\bf 40}, 1371 (1989).


\end{thebibliography}
\end{document}